\documentstyle[prl,aps,multicol]{revtex}
\input epsf

\begin{document}
%\twocolumn
\title{Operation of universal gates in a solid state quantum computer
based on clean Josephson
junctions between d-wave superconductors}
\author{Alexandre Blais$^{a,*}$ and Alexandre M. Zagoskin$^{a,b,c,\dag}$}
\address{$^{a}$D\'epartement de physique and Centre de Recherche en
Physique du Solide, 
Universit\'e de Sherbrooke, Sherbrooke, Qu\'ebec, J1K 2R1, Canada;\\
$^{b}$ DWS Inc., 119-1600 W. 6th Ave., Vancouver, B.C., 
Canada V6J 1R3;\\ 
$^{c}$Physics and Astronomy Dept., U. of British Columbia, 6224
Agricultural Rd., Vancouver, B.C., V6T 1Z1,
 Canada}
\maketitle

\begin{abstract}
The operation of solid state superconducting quantum computer based on
clean Josephson
junctions between two d-wave superconductors is considered.  We show that
freezing of passive
qubits can be achieved using a dynamic global refocusing technique.
Further, we demonstrate
that a universal set of gates can be realized on this system, thereby
proving its universality.
\end{abstract}

\begin{multicols}{2}

Quantum computation algorithms promise of enormous speed up in dealing 
with certain classes of problems\cite{Shor,Grov} can only be realized if
a
quantum computing device is built  on a
scale of at least several thousand qubits. The inherent scalability  of
solid state devices and high level of expertise existing in industrial 
electronics and experimental mesoscopic physics make solid state-based 
quantum computers an attractive choice\cite{divin,Schoen}. The  problem of
quantum coherence preservation in such devices, in the presence  of
macroscopic number of degrees of freedom, is difficult but at least 
theoretically solvable\cite{divin,Schoen}. Moreover, in a recent
experiment 
on a superconducting quantum dot  (single electron transistor, SET)\cite
{Pashkin} coherent quantum beats  were demonstrated in this mesoscopic
system, which proves its suitability  as a qubit prototype.

The coherent ground state and gapped excitation spectrum in
superconductors
make coherence preservation more achievable; there exist already several
suggestions for quantum computers based on Josephson junctions and
superconducting
SETs\cite{Schoen,Gesh,Zag}.

In this paper we consider operation of quantum gates in a solid state
quantum computer based on clean  Josephson junctions 
between d-wave superconductors (i.e. ballistic DND or D-(grain boundary)-D
junctions) \cite{Zag}.
Terminal B of the junction (Fig.1)
is formed by a massive d-wave superconductor; in a multiple-qubit system,
B would be a common ``bus" bar. Terminal A is small enough
to allow,  when isolated, quantum phase fluctuations. It is essentially 
the sign of the superconducting phase difference $\varphi$ between the
terminals A and B that plays the role of ``spin variable" of quantum
computing. The collapse of the wave function is achieved by connecting
terminal A to the equilibrium  electron reservoir (``ground") through a
parity key (superconducting  SET), thus blocking phase fluctuations due
to phase-number uncertainty relation\cite{Tinkham}. Other parity keys,
with different parameters, are used to link adjacent qubits, allowing for
controllable entanglement. (A parity key only passes Cooper pairs, and
only at a certain gate voltage $V_g$\cite{Gisselfalt,Ktototam}.)

\begin{figure}
\epsfxsize=3in
\epsfbox{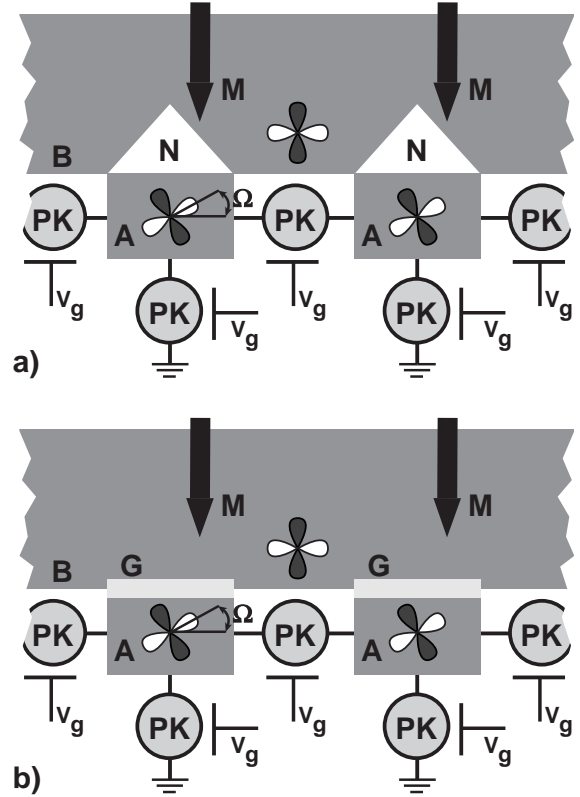}
\caption{a) Superconducting DND qubits: A,B are d-wave superconductors, N
normal conductor, PK  parity key,
 M  scanning tip, $\Omega$ the mismatch angle between the lattices of A
and B. The cut in B is here along $(110)$
 and $(1\bar{1}0)$ directions. Positive lobes of d-wave order parameter
are shaded. Two qubits are shown.
 b) Version of a) using grain boundary (G) junctions.}
\end{figure}

The dynamics of the device was considered in \cite{Zag}. It is
characterized by the phase difference $\varphi$ between terminals A and B, which plays
the role of the position of a quantum particle with mass $M \propto C$,
$C$ being the classical capacitance of the small terminal, in an effective
two-well potential $U(\varphi)$ (Fig.2). It is the crucial advantage of clean DXD
junctions, that the equilibrium phase $\pm\varphi_0$ continuously depends
on the angle between crystal lattices of A and B (and therefore on the d-wave
order parameters in these terminals) in the interval $[0,\pi[$ allowing for
exponentially wide tuning of the tunneling rate\cite{Zag,ZagOsh}.
Moreover,due to time-reversal symmetry breaking in the system, states with $\varphi
= -\varphi_0$ and $\varphi = \varphi_0$ are always degenerate and can be
used as basic $\left|0\right>$ and $\left|1\right>$ states of a qubit\cite{Gesh,Zag}.

\begin{figure}
\epsfxsize=3in
\epsfbox{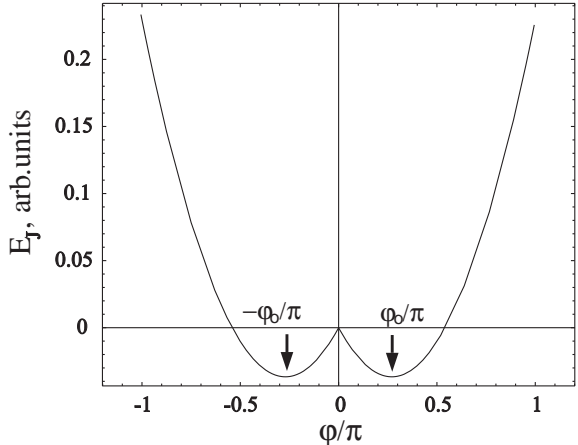}
\caption{Effective potential profile of the system. Minima at
$\pm\varphi_0$ correspond
to "up" and "down" pseudospin states of a qubit. The
mismatch angle is $\Omega=\pi/8$.}
\end{figure}

The basic operations on a qubit are initialization, logical operations
(quantum gates) and measurement. Measurement  is a two-step procedure
and can be performed simultaneously on all qubits or selectively on
individual or groups of qubits. The first step, collapse of the
wave function, is achieved by grounding terminal A. Readout is facilitated
by the existence of small persistent currents and magnetic fluxes
($\ll \Phi _{0}$) which flow in opposite directions in the $|0\rangle $
and $
|1\rangle $ states\cite{Zag,ZagOsh}. While too small to lead to unwanted
inductive coupling between the qubits or decoherence, they can still be
used
to read out the state of the qubit once it was collapsed in one of the
states with $\pm \varphi _{0}$, e.g. using a magnetic force microscope tip
(which is removed during the computations). 
The estimated magnetic moment of order $10^{5}$ to $10^{6}\mu _{B}$ is on
the resolution
limit of commercial magnetic force microscopes. The same property can be
used to initialize individual qubits or whole registers, since this small
coupling to an external field can put the qubit in a desired ($|0\rangle $
or $|1\rangle $) initial state.

Let us now describe how logical operations can be realized in this system. 
In order to maintain coherence, the qubit's electrodes A are isolated from
ground
while performing logical operations. The basic one-qubit logical
operations
are rotations around the $x$ and $z$ axes, $X(\theta)$ and $Z(\phi)$:

\begin{eqnarray}
X(\theta)=e^{-i\sigma_x\theta/2}; \\
Z(\phi)=e^{-i\sigma_z\phi/2}.
\end{eqnarray}

\noindent Operation $X(\theta )$, where $\theta =2t\Delta /\hbar $ and
$\Delta$ are the tunneling matrix elements, is
provided by natural quantum beats between the two basis states $|0\rangle
$ and $|1\rangle $. On the other hand, an effective
rotation around the $z$ axis is realized by lifting the degeneracy of the
basis states by an amount exceeding the tunneling width.
Thereby tunneling between the basis states is suppressed and the natural
oscillations between the basis states, $X(\theta )$, do not interfere with
$Z(\phi )$ operations. The degeneracy between up/down
states can be lifted in various ways. For example,  it can be achieved by
directly applying a localized magnetic field using a magnetic scanning
tip. Other
implementations will be discussed elsewhere. 

As stated, the idle-state of this system corresponds to the logical
operation $X(\theta )$. For a single qubit `computer' this poses no
problem
as logical operations would be applied sequentially without waiting times
(``do-nothing'' periods). In the case that a ``do-nothing'' period is
desired, one can choose this time to be a multiple of the
oscillation period. Thus, using the above convention, this is equivalent
to
applying $X(2n\pi )=\openone$, with $n$ an integer and $\openone$ the
identity operator. The situation with more than one qubit
is less straightforward. Here, we explicitly need passive qubits (qubits 
which undergo no logical operations) to be ``frozen'' during operation on
the
active qubits (qubits over which a logical operation is applied). For
instance, if $Z(\phi )$ is applied on qubit one, the state of passive
qubits must not change during this operation. Since the application
time of logical gates will typically be incommensurate with the time
required for $X(\theta )$ to be equal to $\openone$, a scheme to freeze
passive qubits is
necessary. For this sake, it can be advantageous to have an idle-state
where
the energy of $|0\rangle $ and $|1\rangle $ are degenerate and tunneling
is
coherently blocked. One way to do this would be to temporarily enlarge the
capacitance of electrode A by linking it with an external circuit as it
was suggested by Ioffe {\it et. al.} in their ``quiet-qubit'' proposal
\cite
{Gesh}. However, such an approach brings the risk of losing coherence due
to inelastic processes in
the external normal circuits. On the other hand, making the external
capacitor superconducting would bring uninvited evolution due to Josephson
coupling between the external capacitor and electrode A.

Our suggestion is to employ instead a technique of dynamic global
refocusing
closely related to refocusing methods of NMR \cite{slichter,ernst} and
strong
focusing of accelerators physics \cite{STIFF}. It relies on periodic
perturbation
of the two-well potential with amplitude $\delta E$ slightly exceeding the
tunneling width. In this scheme, the energy shift between the basis states
is periodically varied from $-\delta E$ to $\delta E$.
Explicitly, this corresponds to the pulse sequence

\begin{equation}
\cdots -Z(\delta E\tau/\hbar)-Z(-\delta E\tau/\hbar)-Z(\delta
E\tau/\hbar)-\cdots
\end{equation}

This results in a time dependent angle of rotation around the $z$ axes
which is given, in
the ideal case, by a triangular function of period $2\tau $, the period of
the refocusing
sequence.

The evolution operator for a single qubit is then given, without
approximation, by the Magnus expansion\cite{ernst}~: $
U(t)=\exp\:[-i\sigma_z \int _0^t
dt^{\prime}\phi^{\prime}(t^{\prime})/2]=\exp\:[-i\sigma_z (\delta
E\tau-\delta E\tau+\delta E\tau-\cdots)/2\hbar]$ so that, in the worst
case,
it is equal to $\exp[\pm i\sigma_z \delta E\tau/2\hbar]$. For $\tau$
sufficiently small this reduce to $
U(t)\approx\openone$. Hence, this yields a true idle-state as the
information encoded by the qubits is not perturbed by tunneling nor by
accumulation of relative phase
between the basis states. The characteristic time scale of the refocusing
pulse 
must be much less than the tunneling time (estimated in \cite{Zag} as
$\sim 10^{-8}$ s). 

It was recently demonstrated (Viola and Lloyd\cite{viola:98}
using the spin-boson model; Viola, Knill and Lloyd\cite{viola:99a,viola:99b} under
more general assumptions) that in the limit of {\em very} small 
$\tau$, global refocusing leads to decoherence suppression 
in the $\sigma_x$ and $\sigma_y$ channels (phase decoherence)
provided that $\phi=\pi$ and that delays between the  refocusing pulses are smaller,
or of the order of, the correlation time of the environment \cite{fn2}. 
This correlation time is given by the inverse of a natural cutoff
frequency $\tau_c\sim\omega_c^{-1}$ and determines the fastest time scale of the
environment.  In the case of semiconductor-based structures, where decoherence is
due to phonons, $\tau_c$ is given by the inverse of the Debye frequency
$\omega_c^{-1}\approx 10^{-13}s$ \cite{viola:99a}.
In the present situation, for $\tau$ to be very small requires $\tau\ll
t_{b}$, where $t_{b}\sim l/v_f$ is the ballistic time 
(the time required for the  formation of Andreev levels in the normal part
of the system), $l$ the size of the system and
$v_f$ the Fermi velocity.  Taking $l\sim 10^{3}\AA$ and $v_f\sim
10^{7}cm/s$ \cite{Zag}, we arrive at $\tau\ll 10^{-12}s$,
a similar estimate as in \cite{viola:99a}. Another potentially dangerous
source of decoherence comes from the localized degrees of freedom (nuclear
spins and paramagnetic impurities)\cite{StamP}. The estimates based on the
central spin model\cite{StamP}  show that the relevant energies correspond
to much longer times, in excess of $10^{-8}$ s. (The same estimate can be
made for the decoherence time from these subsystems.) On the other hand,
the dynamics of a spin bath is much more complicated than the one of
oscillator bath or spin boson models, and its behavior under global
refocusing should be a subject of  special investigation.

Logical gates can be performed simultaneously with global refocusing
pulses. Indeed, because the refocusing pulses obviously commute with
$Z(\phi)$,
refocusing can be applied to all qubits (actives and passives)
while performing $Z(\phi)$ on a qubit or in parallel on a group of qubits.
The
evolution of the active qubits is then given by $\exp\:[-i\sigma_z( \int
_0^t dt^{\prime}\phi^{\prime}(t^{\prime})+\phi)/2]~\approx Z(\phi)$. As a
result,
application of $Z(\phi)$ on, e.g. the first qubit, in combination with the
refocusing sequence yields the desired overall action on all qubits~:
$Z(\phi)\otimes\openone\otimes\cdots\otimes
\openone$. On the other hand, applying $X(\theta)$ reduces to stopping
refocusing pulses on the active qubits for a determined period of time.
This  also yields the
desired overall action \cite{note}.

In order to create entangled states, non-local gates are required. Such
an entangling two-qubit operation is realized in this system by opening the
parity key joining two adjacent qubits, Fig 1. With this parity key open,
a Josephson current flows
between states of opposite phases. Thus, the combination $| 00 \rangle$
and 
$| 11 \rangle$ carries no current while $| 01 \rangle$ and $| 10 \rangle$
does. As a result, states of opposite phase will differ from those of
identical phase by a Josephson energy $E_J~ \sim 1-\cos(2\varphi_0)$. The
evolution of a pair of
qubits in this situation then corresponds to a conditional phase shift
($CP$) and,
to an irrelevant phase factor, can be represented in the computational
basis 
$\{~| 00 \rangle,| 01 \rangle,| 10 \rangle,| 11 \rangle~\}$ as

\begin{equation}
CP(\gamma)=\text{Diag}(e^{i\gamma/2},e^{-i\gamma/2},e^{-i\gamma/2},e^{i
\gamma/2}), 
\end{equation}

\noindent with $\gamma~=E_Jt/\hbar$. 
Because $CP(\gamma)$ is diagonal in the computational basis, it commutes
with $Z(\phi)$.
As a result, and under the assumption that the Josephson
energy only weakly perturbs individual two-well potentials\cite{note2},
the latter operation can
be performed simultaneously with the global refocusing
sequence. This condition can always be realized by tuning the gate voltage
on the parity
key, thus varying its transparency and Josephson energy.

Using the three basic operations defined above, it is possible to
construct
a Controlled-Not gate. This operation, denoted $CN_{ij}$ where $i$ and $j$
are the control and target qubits respectively, acts as~: $CN_{12}| i,j
\rangle~=| i,i\oplus j \rangle$, with $\oplus$ denoting addition modulo 2.
Using the above expressions for one- and two-qubits gate, $CN_{12}$ is
realized in this
system, up to an irrelevant global phase factor, by the following sequence

\begin{eqnarray}
CN_{12}= e^{i5\pi/4}X_2(\pi/2)Z_2(\pi/2)X_2(\pi/2)Z_2(\pi/2)\nonumber \\
Z_1(\pi/2)CP(\pi/2) X_2(\pi/2)Z_2(\pi/2)X_2(\pi/2).
\end{eqnarray}

\noindent In this expression, $X_i(\theta)$ ($Z_i(\phi)$) applies $
X(\theta)$ ($Z(\phi)$) on the $i^{th}$ qubit while leaving the others
unchanged (e.g.,
$Z_1(\phi)~=Z(\phi)\otimes\openone\otimes\cdots\otimes\openone$).

In the setup of figure 1, it is possible to apply two-qubit gates only to
adjacent qubits. It is therefore necessary to introduce a swap operator,
denoted $SW_{ij}$, which exchanges the states of qubits $i$ and $j$. A
swap on two
adjacent qubits is realized by the following combination of Controlled-Not
gates

\begin{equation}
SW_{12}=CN_{12}CN_{21}CN_{12}.
\end{equation}

\noindent Using this operator repeatedly, it is then possible to juxtapose
any chosen pairs of qubits and, as a result, to apply Controlled-Not gates
on any chosen pairs of qubits.

Because of the commutation relations between the Pauli operators,
combinations of rotations around the $x$ and $z$
axes generate $SU(2)$, the group of 2 by 2 unitary matrices with
determinant +1.  Thus, it is possible to realize all
one-qubit gates on this system. Furthermore, as been shown by Barenco {\it
et. al.},
the set of all single qubit gates and the Controlled-Not is complete for
quantum computation\cite{barenco:95}. It is therefore possible to generate
all of $
SU(2^{n})$ with proper sequences of gates in such a $n$-qubit DXD
superconducting quantum computer. 

In conclusion, we have shown that a solid state superconducting quantum
computer
suggested in \cite{Zag} allows application of a complete set of quantum
logical gates and is therefore a realization of a universal quantum
computer.

\vskip 0.3cm
\noindent {\bf Acknowledgements} We are grateful to Martin Beaudry and
Philip Stamp for helpful discussions and particularly to
Serge Lacelle and Andr\'e-Marie Tremblay for stimulating discussions and a
critical reading of the manuscript.  A.B. received support from FCAR. A.Z.
was partially supported by FCAR and CIAR.

\vskip 0.7cm
\noindent ${}^*$ Electronic address: ablais@physique.usherb.ca\\ 
${}^\dag$ Electronic address: zagoskin@physics.ubc.ca\\

\end{multicols}


\begin{references}
\bibitem{fn2} Here $\sigma_{x,y}$-channels refer
 to the processes described by the corresponding Pauli matrices in the 
computational basis and involving coherent tunneling between its
states (cf. Eq.(1)).
\bibitem{note} Freezing of passive qubits during application of $Z(\phi)$
is unnecessary if the characteristic time of the $z$-rotation is
much faster than the tunneling time. On the contrary, freezing cannot be
avoided during $X(\theta)$ operations whatever its characteristic time
compared to $Z(\phi)$.
\bibitem{note2} The critical Josephson current through the open parity key
must be thus
chosen to be  much less than the one through the clean d-d boundary.
\end{references}
\end{document}